\documentclass[prx,twocolumn,superscriptaddress,showpacs,amsmath,amssymb,amsxtra]{revtex4-1}
\usepackage[dvips]{graphicx}
\usepackage{subfigure}
\usepackage{natbib}
\usepackage{bm,array}

\def\U#1{{\rm #1}} 
\def\u#1{_{\rm #1}}
\newcommand{\ket}[1]{| #1 \rangle}

\newcommand{\expect}[1]{{\langle #1 \rangle}} 
\newcommand{\normal}[1]{{\langle: #1 :\rangle}} 

\def\Dsa{\U{D}\u{s1}}
\def\Dsb{\U{D}\u{s2}}
\def\Dsab{\U{D}\u{s1(2)}}
\def\Dva{\U{D}\u{v1}}
\def\Dvb{\U{D}\u{v2}}
\def\Dvab{\U{D}\u{v1(2)}}
\def\Dta{\U{D}\u{t1}}
\def\Dtb{\U{D}\u{t2}}
\def\Dtab{\U{D}\u{t1(2)}}
\def\Da{\U{D}\u{3}}
\def\Db{\U{D}\u{4}}
\def\g2{g^{(2)}}

\newcolumntype{P}[1]{>{\centering\arraybackslash}p{#1}}

\begin{document}
\title{
Heralded single excitation of atomic ensemble 
via solid-state-based telecom photon detection
}
\author{Rikizo~Ikuta}
\affiliation{Graduate School of Engineering Science, Osaka University,
Toyonaka, Osaka 560-8531, Japan}
\author{Toshiki~Kobayashi}
\affiliation{Graduate School of Engineering Science, Osaka University,
Toyonaka, Osaka 560-8531, Japan}
\author{Kenichiro~Matsuki}
\affiliation{Graduate School of Engineering Science, Osaka University,
Toyonaka, Osaka 560-8531, Japan}
\author{Shigehito~Miki}
\affiliation{Advanced ICT Research Institute, 
National Institute of Information and Communications Technology (NICT),
Kobe 651-2492, Japan}
\author{Taro~Yamashita}
\affiliation{Advanced ICT Research Institute, 
National Institute of Information and Communications Technology (NICT),
Kobe 651-2492, Japan}
\author{Hirotaka~Terai}
\affiliation{Advanced ICT Research Institute, 
National Institute of Information and Communications Technology (NICT),
Kobe 651-2492, Japan}
\author{Takashi~Yamamoto}
\affiliation{Graduate School of Engineering Science, Osaka University,
Toyonaka, Osaka 560-8531, Japan}
\author{Masato~Koashi}
\affiliation{Photon Science Center, Graduate School of Engineering,
The University of Tokyo, Bunkyo-ku, Tokyo 113-8656, Japan}
\author{Tetsuya~Mukai}
\affiliation {NTT Basic Research Laboratories, 
NTT Corporation, Atsugi, Kanagawa 243-0198, Japan}
\author{Nobuyuki~Imoto}
\affiliation{Graduate School of Engineering Science, Osaka University,
Toyonaka, Osaka 560-8531, Japan}

\begin{abstract}
Telecom photonic quantum networks with matter quantum systems 
enable a rich variety of applications, 
such as a long distance quantum cryptography and one-way quantum computing. 
Preparation of a heralded single excitation~(HSE) in an atomic ensemble 
by detecting a telecom wavelength photon having a correlation with the atomic excitation 
is an important step. 
Such a system has been demonstrated with a quantum frequency conversion~(QFC) 
to telecom wavelength employing a Rb atomic cloud. 
However the limited wavelength selection prevents the next step. 
Here we for the first time demonstrate HSE with a solid-state-based QFC 
and a detector for a telecom wavelength 
that will have a great advantage of the utility of mature telecom technologies. 
We unambiguously show that the demonstrated HSE indicates 
a non-classical statistics by the direct measurement of the autocorrelation function. 
\end{abstract}

\maketitle

\section{introduction}
\begin{figure}[t]
 \begin{center}
  \scalebox{0.34}{\includegraphics{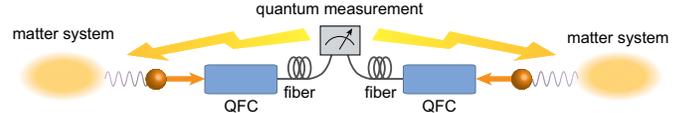}}
  \caption{
  Concept of an elementary link of quantum matter systems 
  through the optical fiber quantum communication by telecom photons. 
  The telecom photons converted from light pulses which have the correlation 
  with the matter systems are measured 
  for heralding the connection between the quantum matter systems. 
  \label{fig:concept}}
 \end{center}
\end{figure}

Recent advances in telecom photonic quantum information technology 
allow integrated photonic circuits, 
ultra fast switching, and highly efficient single photon detection 
and wavelength conversion. 
Combining those technologies with matter quantum systems 
as shown in Fig.~\ref{fig:concept}  
will 
certainly open up a new avenue for the advanced quantum information technologies, 
such as a long-distance quantum key distribution~\cite{Ekert1991,Lo2014}, 
quantum computation including measurement-based topological methods~\cite{Ladd2010,Morimae2012}, 
and quantum communication among separated nodes~\cite{Kimble2008}, 
and for fundamental tests of the physics outside of the light cone~\cite{Hensen2015}. 
Manipulation of a quantum state in matter quantum systems has been performed 
by various wavelengths of 
the photons~\cite{Yuan2008,Olmschenk2009,Ritter2012,Hofmann2012,Bernien2013,Delteil2015}, 
but the photons in the demonstrations are of visible or near-infrared wavelengths 
which are not compatible with the telecom regime. 
Thus to fill the wavelength mismatch, 
the quantum frequency conversion~(QFC) 
which preserves non-classical photon statistics 
and a quantum state of an input light has been studied~\cite{Kumar1990}. 

A pioneering work for this task is a QFC based on a four wave mixing 
operated around near resonance in cold rubidium~(Rb) atoms~\cite{Radnaev2010,Dudin2010}. 
In this experiment, a visible photon which heralds spin-wave excitations 
in cold rubidium~(Rb) atoms as a quantum memory 
was frequency-down-converted to a telecom photon by the QFC by the atomic cloud. 
The QFC device 
has a high efficiency due to the operation near atomic resonance. 
In the meanwhile, it limits choice of wavelengths available to the QFC. 
Recently widely tunable solid-state-based QFCs have been demonstrated 
by using various non-linear optical crystals\cite{Tanzilli2005,Rakher2010,Ikuta2011,Zaske2012,Kobayashi2016}, 
and the potential ability for the single photon source 
with a cold Rb atom ensemble has been demonstrated~\cite{Albrecht2014}. 
However, 
a heralded single excitation~(HSE) 
in long-lived matter quantum system by telecom photon detection 
required for quantum network illustrated in Fig.~\ref{fig:concept} 
has not been demonstrated. 

We first show an HSE in a cold Rb atomic ensemble 
through the detection of the photon emission from $\U{D}_2$ line of Rb. 
Subsequently, we show the conversion of the wavelength of the photon to 1522~nm 
in the telecom range by a QFC 
using a periodically-poled lithium niobate~(PPLN) waveguide. 
Finally we show the demonstration of the HSE 
through the telecom photon detection by 
superconducting single photon detectors~(SSPDs). 
Remarkably, we show the non-classical property 
of the HSE by observing the autocorrelation functions 
in addition to the cross correlation functions. 
Furthermore, we show that 
the telecom photons heralded by the atomic spin state 
also has a non-classical photon statistics. 

\section{Experimental setup}
\begin{figure*}[t]
 \begin{center}
 \scalebox{0.4}{\includegraphics{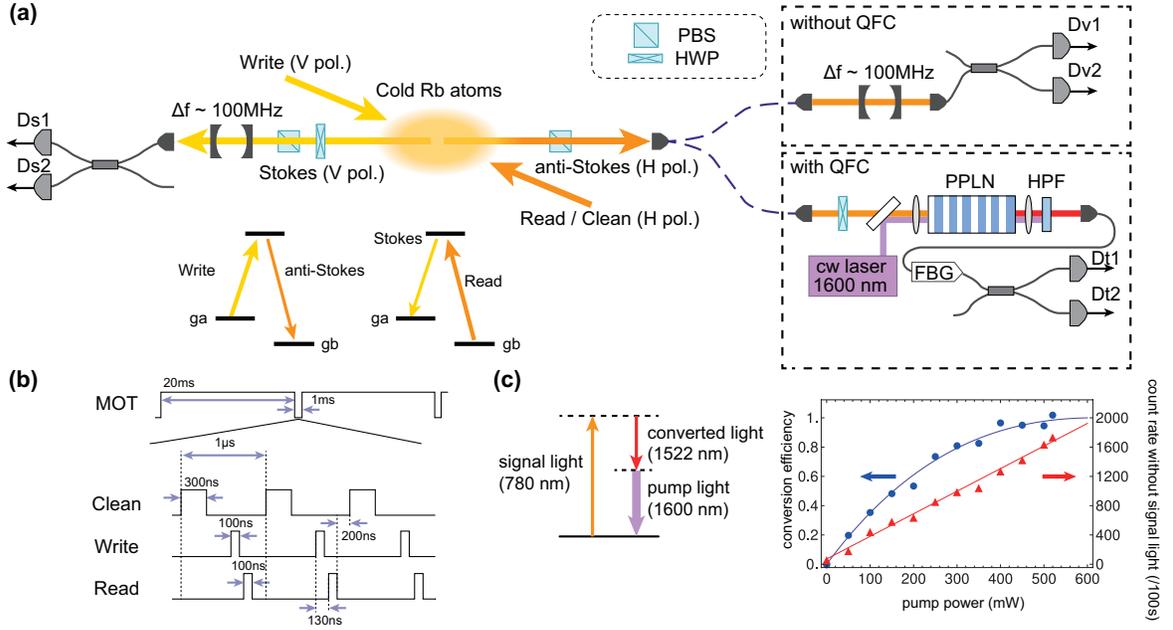}}
  \caption{
  (a) 
  Our experimental setup without and with the QFC, 
  and the $\Lambda$-type energy levels 
  of $\U{D}_2$ line in $^{87}\U{Rb}$ with the relation of the light pulses 
  used in our experiment. 
  The strong write pulse is removed from the optical path of the anti-Stokes photons 
  by the spatial, polarizing and the frequency separation. 
  The filtering of the write pulse from the Stokes photons 
  is performed by the large difference of 
  the spatial direction and the temporal separation. 
  The separation of the read light from the signal photons is 
  performed in the same manner. 
  (b)
  Time sequence of the experiment. 
  (c)
  The left figure is the relevant energy levels of QFC. 
  The difference frequency generation of a photon 
  from 780~nm to 1522~nm is performed by using a strong pump light at 1600~nm. 
  The right graph shows the conversion efficiency 
  and the noise counts induced by the frequency converter. 
  \label{fig:setup}}
 \end{center}
\end{figure*}
The $^{87}\U{Rb}$ atomic ensemble 
is prepared by a magneto-optical trap~(MOT) for 20~ms. 
After the lasers and the magnetic field for the MOT are turned off, 
we use $\Lambda$-type energy levels of $\U{D}_2$ line at 780~nm 
($5^{2}S_{1/2}\leftrightarrow 5^{2}P_{3/2}$) for the experiment 
as shown in Fig.~\ref{fig:setup}~(a). 
Two of the ground levels having  
$F=2$ and $F=1$ are denoted by $g\u{a}$ and $g\u{b}$, respectively, 
in which magnetic sublevels are degenerated 
because of the absence of the magnetic field. 
At first, a horizontally~(H-) polarized 300-ns clean pulse 
at the resonant frequency between $g\u{b}$ and the excited level~($F'=2$) 
prepares the atomic ensemble into $g\u{a}$. 
Then a vertically(V-) polarized 100-ns write pulse 
near the resonant frequency between $g\u{a}$ and the excited level 
provides the Raman transition from $g\u{a}$ to $g\u{b}$ 
and anti-Stokes~(AS) photons simultaneously. 
In our experiment, the AS photons emitted in a direction at a small angle~($\sim 3^\circ$) 
relative to the direction of the write pulse is detected with H polarization. 
We will explain an optical circuit for the AS photons in detail later. 
The photon detection in mode AS tells us that 
the single excitation has been prepared in the atomic ensemble, 
which we call the heralded single excitation~(HSE). 
In order to read out the HSE, 
an H-polarized 100-ns read light at the resonant frequency 
between $g\u{b}$ and the excited level is injected to the atoms. 
The Raman transition to $g\u{a}$ provides Stokes~(S) photons. 
By setting the direction of the read pulse 
to be opposite to the write pulse, 
the Stokes photons are emitted in a mode~(S) traveling in the opposite direction 
to that of mode AS 
due to the phase matching condition of the four light pulses~\cite{Chen2007}. 
In our experiment, we detect the photons in mode S 
with V polarization only. 
The S photons pass through 
a cavity-based bandpass filter~\cite{Palittapongarnpim2012} 
with a bandwidth of $\sim$100~MHz, 
and then they are coupled to a single-mode optical fiber. 
The photons are divided into two by a fiber-based half beamsplitter~(HBS), 
and then they are detected
by silicon avalanche photon detectors~(APDs) denoted by $\Dsa$ and $\Dsb$. 

Let us explain the optical circuit for the photons in mode AS. 
The H-polarized AS photons from the atomic ensemble 
is coupled to a polarization maintaining fiber~(PMF). 
Without QFC, the PMF is connected to an optical circuit 
for the AS photons passing through a cavity-based bandpass filter 
with a bandwidth of $\sim$100~MHz. 
After the filtering, 
the AS photons are divided into two by a HBS. 
Then they are detected by using APDs denoted by $\Dva$ and $\Dvb$. 
With QFC, 
the PMF is connected to the optical circuit for the QFC. 
The QFC converts the wavelength of the AS photons 
from 780~nm to 1522~nm~\cite{Ikuta2011,Ikuta2013}. 
The conversion efficiency 
and the background-noise rate induced by the 1600-nm pump light for the conversion 
with respect to the pump power are shown in Fig.~\ref{fig:setup}~(c). 
The details of the QFC device are shown in Appendix~\ref{appA}. 

After the QFC, 
the 1522-nm photons are coupled to a single-mode optical fiber. 
The 1522-nm photons pass through a fiber Bragg grating~(FBG) 
with a bandwidth of $\sim $1~GHz followed by a HBS. 
Finally, the 1522-nm photons divided by the HBS 
are detected by SSPDs~\cite{Miki2013,Yamashita2013} 
denoted by $\Dta$ and $\Dtb$. 

\begin{figure}[t]
 \begin{center}
 \scalebox{0.55}{\includegraphics{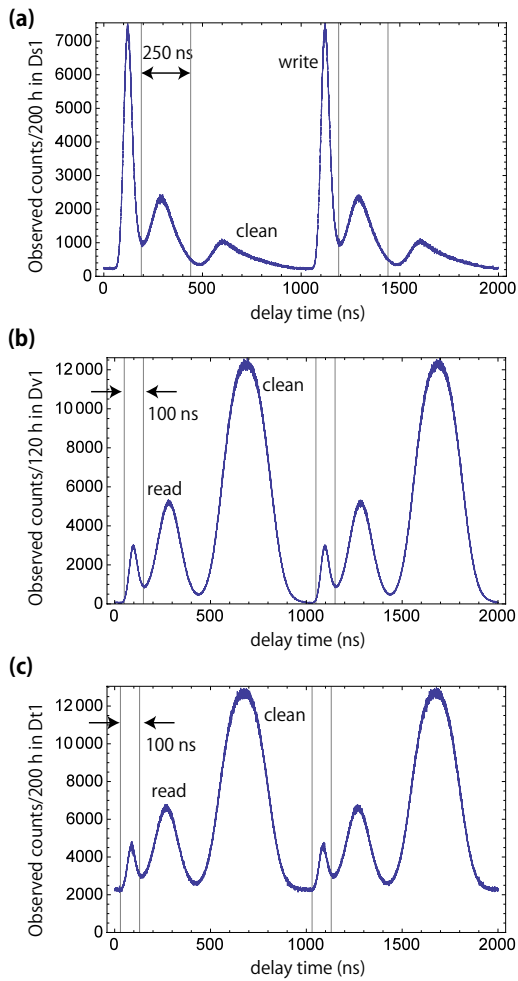}}
  \caption{
  The observed stop signals by (a) $\Dsa$, (b) $\Dva$ and (c) $\Dta$. 
  In every figures, 
  the interval of the largest signals is 1~$\mu$s. 
  \label{fig:histogram}}
 \end{center}
\end{figure}
The period of the write pulses is 1~$\mu$s, 
as shown in Fig.~\ref{fig:setup}~(b). 
The injection of the write pulses is repeated 990 times 
until the MOT is turned on again. 
In each sequence, 
we collect coincidence events 
between S and AS photon detections 
by using a time-digital converter~(TDC). 
A start signal of TDC is synchronized to the timing 
of triggering the write pulse. 
All electric signals from photon detections, 
i.e. $\Dsa$, $\Dsb$, $\Dva$, $\Dvb$, $\Dta$ and $\Dtb$ can be used 
as stop signals for TDC. 
Typical histograms of the coincidence events in 1 sequence 
are shown in Figs.~\ref{fig:histogram}~(a)--(c). 
In the following experiments, 
we postselect 
the coincidence events within the 250-ns and 100-ns time windows 
for the signals of modes S and AS, respectively. 
In the histograms, while several unwanted peaks 
coming from the write, read and clean pulses are observed in 1 sequence, 
they are temporally separated from the main signals and 
can be eliminated by the selection of the time windows. 

\section{Experimental results}
\subsection{Cross correlation function vs. the write power} 
\begin{figure}[t]
 \begin{center}
 \scalebox{0.5}{\includegraphics{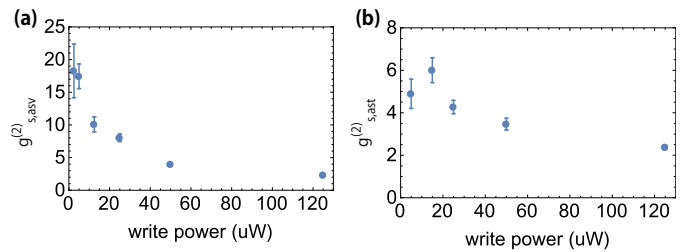}}
  \caption{
  The dependencies of the cross correlation functions 
  on the power of the write pulse. 
  (a) $\g2\u{s,asv}$ vs write power without the QFC, 
  and 
  (b) $\g2\u{s,ast}$ vs write power with the QFC. 
  \label{fig:cross}}
 \end{center}
\end{figure}
As a preliminary experiment, 
we measured the cross correlation function 
between the Stokes and the anti-Stokes photons 
with and without QFC 
for various powers of the write pulse. 
We denote the detection probabilities of the detectors 
$\Dsab$, $\Dvab$ and $\Dtab$ 
by $p\u{s1(2)}$, $p\u{v1(2)}$ and $p\u{t1(2)}$, respectively. 
We derive the cross correlation functions by the equations 
$\g2\u{s,asv}=p\u{s,v}/(p\u{s}p\u{v})$ 
without the QFC 
and 
$\g2\u{s,ast}=p\u{s,t}/(p\u{s}p\u{t})$ 
with the QFC, 
where $p_i$ is a probability of the photon detection 
by $\U{D}_{i1}$ or $\U{D}_{i2}$ for $i=\U{s,\, v,\, t}$, 
and 
$p\u{s,v}$ and $p\u{s,t}$ are coincidence probabilities 
between ($\Dsa$ or $\Dsb$) \& ($\Dva$ or $\Dvb$), 
and ($\Dsa$ or $\Dsb$) \& ($\Dta$ or $\Dtb$), respectively. 
The experimental results are shown in Fig.~\ref{fig:cross}. 
From Fig.~\ref{fig:cross}~(a), without the QFC, 
we see that 
the AS photons which are well correlated with the Stokes photons were prepared 
for a small power of the write pulse. 
Fig.~\ref{fig:cross}~(b) shows 
the cross correlation function when the QFC was performed. 
We see that the correlation is still kept after the QFC of the AS photons, 
while its value declines. 
From the experimental results in Figs.~\ref{fig:cross}~(a) and (b), 
we set the power of the write pulse to $\sim 15\mu$W 
in the following experiments. 

\subsection{Correlation functions without QFC}
Without the QFC, 
we performed the measurement with an integration measurement time of about 120 hours. 
The observed cross correlation function was $\g2\u{s,asv}=9.69\pm 0.04$. 
The autocorrelation function $\g2\u{s,s}$ of the photons in mode S 
without the heralding signal of the photon detection of mode AS was observed 
by the Hanbury-Brown and Twiss setup~\cite{Hanbury1956} using $\Dsa$ and $\Dsb$. 
It is defined by $\g2\u{s,s}=p\u{s1,s2}/(p\u{s1}p\u{s2})$, 
where $p_{\U{s}j}$ is a probability of the photon detection 
by $\U{D}_{\U{s}j}$ for $j=1,2$, 
and 
$p\u{s1,s2}$ is a coincidence probability between $\Dsa$ and $\Dsb$. 
The observed value was $\g2\u{s,s}=1.58\pm 0.03$. 
Similarly, the autocorrelation function of the AS photons 
without the heralding signal of the photon detection of mode S 
measured by $\Dva$ and $\Dvb$ 
was $\g2\u{asv,asv}=1.99\pm 0.03$. 
The autocorrelation function of the S photons 
heralded by the photon detection of mode AS 
is denoted by 
$\g2\u{s,s|asv}=p\u{s1,s2,v}p\u{v}/(p\u{s1,v}p\u{s2,v})$, 
where 
$p_{\U{s}j,\U{v}}$ is a two-fold coincidence probability 
between $\U{D}_{\U{s}j}$ and ($\U{D}\u{v1}$ or $\U{D}\u{v2}$) 
for $j=1,2$, and 
$p\u{s1,s2,v}$ is a three-fold coincidence probability 
among $\U{D}\u{s1}$, $\U{D}\u{s2}$ and ($\U{D}\u{v1}$ or $\U{D}\u{v2}$). 
The observed autocorrelation function was 
$\g2\u{s,s|asv}=0.34\pm 0.05$. 
In the case where AS photons heralded by S photons, 
the autocorrelation function of the heralded AS photons was 
$\g2\u{asv,asv|s}=0.47\pm 0.06$. 

\subsection{Correlation functions with QFC} 
With the QFC, 
we performed the experiment with an integration time of about 200 hours. 
The observed cross correlation function 
between the S and the 1522-nm photons 
converted from the AS photons was $\g2\u{s,ast}=4.09\pm 0.01$. 
The autocorrelation function of the 1522-nm photons 
without the heralding signal of the S photons was measured by 
using $\Dta$ and $\Dtb$, and the observed value was 
$\g2\u{ast,ast}=1.12\pm 0.01$. 
The autocorrelation function of the S photons 
heralded by the 1522-nm photons converted from the AS photons 
was $\g2\u{s,s|ast}=0.71\pm 0.10$. 
The value remains smaller than 1 after the QFC, 
and thus we conclude that the demonstrated HSE in Rb atoms 
shows non-classical statistics 
with the detection of the 1522-nm photons. 
In addition, 
the autocorrelation function of the 1522-nm photons 
with the heralding signal by the S photons 
was observed to be $\g2\u{ast,ast|s}=0.54\pm 0.09$. 
In Table~\ref{tbl:view}, 
we list all the observed correlation functions. 
Especially, autocorrelation functions having less than 1 
unambiguously show the non-classical statistics. 

\begin{table}[t]
\begin{center}
\begin{tabular*}{1\linewidth}{@{\extracolsep{\fill}}cccccccccccc}
\hline
\multicolumn{12}{|c|}{cross correlation functions} \\ \hline
\multicolumn{6}{c}{$\g2\u{s,asv}$} & 
\multicolumn{6}{c}{$\g2\u{s,ast}$} \\ 
\multicolumn{6}{c}{$9.69\pm 0.04$} & 
\multicolumn{6}{c}{$4.09\pm 0.01$} \\ \hline

\multicolumn{12}{|c|}{autocorrelation functions without heralding signals} \\ \hline

\multicolumn{4}{c}{$\g2\u{s,s}$} & 
\multicolumn{4}{c}{$\g2\u{asv,asv}$} &
\multicolumn{4}{c}{$\g2\u{ast,ast}$} \\ 

\multicolumn{4}{c}{$1.58\pm 0.03$} & 
\multicolumn{4}{c}{$1.99\pm 0.03$} &
\multicolumn{4}{c}{$1.12\pm 0.01$} \\ \hline

\multicolumn{12}{|c|}{autocorrelation functions with heralding signals}\\ \hline

\multicolumn{3}{c}{$\g2\u{s,s|asv}$} & 
\multicolumn{3}{c}{$\g2\u{s,s|ast}$} & 
\multicolumn{3}{c}{$\g2\u{asv,asv|s}$} & 
\multicolumn{3}{c}{$\g2\u{ast,ast|s}$} \\ 

\multicolumn{3}{c}{$0.34\pm 0.05$} & 
\multicolumn{3}{c}{$0.71\pm 0.10$} & 
\multicolumn{3}{c}{$0.47\pm 0.06$} & 
\multicolumn{3}{c}{$0.54\pm 0.09$} 
\end{tabular*}
 \caption{
 The observed values of the correlation functions. 
 \label{tbl:view}}
 \end{center}
\end{table}
From $\g2\u{ast,ast|s}=0.54$, 
we can estimate the visibility of Hong-Ou-Mandel~(HOM) interference 
between two independently prepared telecom photons, 
each of which comes from the atomic ensemble demonstrated here. 
The visibility of the HOM interference is described 
by $V=1/(1+\g2\u{ast,ast|s})$ 
with assumptions that mode matching of the signal photons is perfect 
and stray photons do not exist~(see Appendix~\ref{appB}). 
The estimated value is $V=0.65$, which exceeds the classical limit of 0.5. 

\section{Discussion}
In the following, 
we discuss the degradation of 
the observed autocorrelation functions 
with QFC~($\g2\u{s,s|ast}$ and $\g2\u{ast,ast|s}$) 
with respect to those without QFC~($\g2\u{s,s|asv}$ and $\g2\u{asv,asv|s}$). 
Because the APDs and the SSPDs have very small dark counts, 
the observed cross correlation functions 
and the autocorrelation functions in Table~\ref{tbl:view} are considered as 
intrinsic values of the measured photons. 
Therefore the degradations from $\g2\u{s,s|asv}$ to $\g2\u{s,s|ast}$ 
and from $\g2\u{asv,asv|s}$ to $\g2\u{ast,ast|s}$ are mainly caused 
by the background noise of QFC process of the AS photons. 
We assume that the background noise is induced by 
Raman scattering of the QFC pump light 
which is statistically independent on the signal photons. 

Using the definition of the cross correlation functions  
$\g2\u{s,asv}$ and $\g2\u{s,ast}$, and ratio $\zeta$ 
of the average photon number of the signal in mode AS to the equivalent input noise 
to the converter, we obtain~(see Appendix~\ref{appC}) 
\begin{eqnarray}
\g2\u{s,ast}=\frac{\g2\u{s,asv} \zeta + 1}{\zeta + 1}.
\label{eq:g2cross}
\end{eqnarray}
Furthermore $\g2\u{ast,ast}$ is described by
\begin{eqnarray}
\g2\u{ast,ast}= \frac{1}{(1+\zeta)^2} 
\left( 
\zeta^2 \g2\u{asv,asv} + \g2\u{noise} + 2 \zeta
\right), 
\label{eq:g2auto}
\end{eqnarray}
where $\g2\u{noise}$ is the autocorrelation function of the noise photons~\cite{Ikuta2014}. 
From the observed values~($\g2\u{s,asv}$, $\g2\u{s,ast}$, $\g2\u{asv,asv}$ and $\g2\u{ast,ast}$) 
and Eqs.~(\ref{eq:g2cross}) and (\ref{eq:g2auto}), 
we obtain $\zeta=0.55$ and $\g2\u{noise}=0.99$. 
In Eq.~(\ref{eq:g2auto}), 
replacing 
$\g2\u{ast,ast}$, $\zeta$ and $\g2\u{asv,asv}$ 
by $\g2\u{ast,ast|s}$, $\zeta'=\g2\u{s,asv} \zeta$ and $\g2\u{asv,asv|s}$, 
we have $\g2\u{ast,ast|s} = 0.62$, 
which is in good agreement with the observed value 
within the margin of error. 

In addition, 
$\g2\u{s,s|ast}$ is described by using the observed values 
as~(see Appendix~\ref{appC}) 
\begin{eqnarray}
\g2\u{s,s|ast} 
&=& \g2\u{s,s|asv}\left(
\frac{\g2\u{s,asv}}{\g2\u{s,ast}}\right)^2
\frac{\zeta}{\zeta +1}\nonumber\\
&&
\hspace{1cm}
+ \g2\u{s,s}
\left(
\frac{1}{\g2\u{s,ast}}
\right)^2
\frac{1}{\zeta +1}.
\label{eq:SM}
\end{eqnarray}
Thus, from the observed values~($\g2\u{s,s|asv}$, $\g2\u{s,asv}$ and $\g2\u{s,ast}$), 
and the estimated value of $\zeta=0.55$, we obtain 
$\g2\u{s,s|ast}=0.74$, 
which is also in good agreement with the experiment. 

Below, based on the above estimations, 
we discuss a possible improvement of 
$\g2\u{s,s|ast}$ and $\g2\u{ast,ast|s}$ 
achieved by increasing the value of $\zeta'=\g2\u{s,asv} \zeta$ 
corresponding to the ratio of the average photon number of the signal in mode AS 
to the noise induced by QFC 
conditioned on the photon detection in mode S. 
For this, we discuss three possible methods 
without changing the performance of QFC as follows: 
(a) increase of the collection efficiency of the AS photons, 
(b) proper selection of the polarization of the AS photons, and 
(c) refinement of the experimental system 
for improvement of the correlation functions without QFC. 
(a) 
In our experiment, the collection efficiency of the AS photons without QFC 
is roughly estimated to be several percent~(see Appendix~\ref{appD}). 
Here we borrow the collection efficiency of the AS photons
from the current state-of-the-art experiments~\cite{Laurat2006,Albrecht2014}, 
where the value is about ten times larger 
than that in our experiment, resulting in $\zeta \rightarrow 10\zeta$. 
This implies that the values of the autocorrelation correlation functions will 
be $\g2\u{s,s|ast}=0.39$ and $\g2\u{ast,ast|s}=0.49$. 
In this regime, the QFC preserves almost the same statistics 
as before the conversion shown in Table~\ref{tbl:view}.  
(b) 
In our experiment, 
the magnetic sublevels of the ground states $g\u{a}$ and $g\u{b}$ 
are degenerated, which results in the loss of the AS photons 
by the polarization selection. 
Such signal loss is estimated to be about 0.2 under an assumption 
that the atoms are uniformly distributed in the magnetic sublevels 
as the initial state~(see Appendix~\ref{appE}). 
So its improvement will contribute to increase $\zeta$ by a factor of $1.25$, 
resulting in $\g2\u{s,s|ast}=0.68$ and $\g2\u{ast,ast|s}=0.60$ 
from $\g2\u{s,s|ast}=0.74$ and $\g2\u{ast,ast|s}=0.62$. 
(c) Refinement of the measurement for the S photons 
will increase $\zeta'$ through the improvement of $\g2\u{s,asv}$ 
and other correlation functions without QFC, while $\zeta$ is kept. 
In addition, the use of a smaller excitation probability 
will also be effective for increasing $\g2\u{s,asv}$ 
as shown in Fig.~\ref{fig:cross}~(a). 
While it may need a long time experiment, 
higher collection probabilities of modes AS and S 
discussed in (a) may enable us to 
compensate the reduction of the excitation probability. 

\section{Conclusion}
In conclusion, we have clearly shown 
the HSE in a cold Rb atomic ensemble 
by detection of the photons at the telecom wavelength, 
which has been observed by the direct measurement of the autocorrelation function. 
It was achieved 
by using the solid-state-based QFC and the detectors 
with the high efficiency and low noise properties. 
In addition, we have observed the non-classical photon statistics 
of the converted telecom photons. 
It indicates that observation of a non-classical interference 
between the two telecom photons prepared by duplicating the system 
demonstrated here will be possible. 
The quantum system composed of the matter systems and the telecom photons 
with the solid-state-based QFC and detectors will be useful 
for connecting various kinds of matter systems through mature telecom technology. 

\section*{Acknowledgements}
This work was supported by 
MEXT/JSPS KAKENHI Grant Number 
26286068, 25247068, 15H03704, 16H02214, 16K17772, 
and JSPS Grant-in-Aid for JSPS Fellows 14J04677.

\appendix

\section{Frequency conversion device}
\label{appA}

The QFC is based on difference frequency generation~(DFG) 
by using a periodically-poled LiNbO$_3$~(PPLN) waveguide. 
For the DFG of the signal photon at 780 nm, 
a V-polarized cw pump laser at 1600 nm with a linewidth of 150 kHz is used. 
The pump light is combined with the signal photons at a dichroic mirror. 
Then, they are focused on the type-0 quasi-phase-matched 
PPLN waveguide. 
The length of the PPLN crystal is 20 mm, 
and the acceptable bandwidth is about 0.3 nm 
which is much wider than that of the signal photons 
emitted from the atomic ensemble. 
After passing through the PPLN waveguide, 
the strong pump light is removed 
by a high-pass filter and a bandpass filter with a bandwidth of 1~nm. 
In the measurement of the correlation functions with QFC, 
we set the pump power to be $\sim 200$ mW. 

\section{Visibility of the HOM interference}
\label{appB}

Here we consider the HOM interference 
between two independently prepared signal photons in spatial modes 1 and 2 
without any stray photons. 
The two signal photons are mixed by a HBS, 
and then the output photons in spatial modes 3 and 4 
are detected by detectors $\Da$ and $\Db$ 
with sufficiently wide detection windows. 
The visibility of the HOM interference is defined by 
$V=1-P_0/P_\infty$. 
Here $P_0$ and $P_\infty$ are two-fold coincidence probabilities 
with $\Delta t=0$ and $\Delta t=T$, respectively, 
where $T$ is much larger than the pulse duration of input photons. 
We assume that when $\Delta t=0$, 
mode matching of the input photons is perfect. 
We also assume that the two input photons have 
the same values $\g2_1$ and $s_1$ of the autocorrelation function 
and the average photon number, respectively, 
and are statistically independent. 
We define that the number operator of spatial mode $i(=1,2,3,4)$ is $n_{i}$. 
By using the normal-ordered product of the operators and 
bosonic commutation relations, 
we obtain 
$P_0/\eta=\normal{n_3n_4}=(\normal{n_1^2}+\normal{n_2^2})/4 
=s_1^2\g2_1/2$, 
and 
$P_\infty/\eta=\normal{n_3n_4}=(\normal{n_1^2}+\normal{n_2^2}+2\normal{n_1n_2})/4
=s_1^2(\g2_1+1)/2$, 
where $\eta$ is a product of the quantum efficiencies of the detectors. 
As a result, 
$V=1/(1+\g2_1)$ is obtained. 

\section{Derivation of 
Eqs.~(\ref{eq:g2cross}) and (\ref{eq:SM}) }
\label{appC}

We derive Eqs.~(\ref{eq:g2cross}) and (\ref{eq:SM}) in the main text. 
We define that the number operators 
for mode S, mode AS just before the QFC 
and the equivalent input noise to the frequency conversion device 
are $n\u{s}$, $n\u{asv}$ and $n\u{noise}$, respectively. 
By using the definitions and 
the normal-ordered product of the operators, 
we have 
$\zeta = \expect{n\u{asv}}/ \expect{n\u{noise}}$, 
$g\u{s,asv}^{(2)}=\normal{n\u{asv}n\u{s}}/(\expect{n\u{s}}\expect{n\u{asv}})$, 
$g\u{s,s}^{(2)}=\normal{n\u{s}^2}/\expect{n\u{s}}^2$ 
and 
$g\u{s,s|asv}^{(2)}=\normal{n\u{s}^2n\u{asv}}\expect{n\u{asv}}/(\normal{n\u{s}n\u{asv}}^2)$. 
Since the noise photons and the S photons are statistically independent, 
$\normal{n\u{s}n\u{noise}}=\expect{n\u{s}}\expect{n\u{noise}}$ 
and 
$\normal{n\u{s}^2n\u{noise}}=\expect{n\u{s}^2}\expect{n\u{noise}}$ 
are satisfied. 
The number operator $n\u{ast}$ for the mode of the 1522-nm photons 
converted from the AS photons is described by 
$n\u{ast}=\eta\u{conv} (n\u{asv}+n\u{noise})$, 
where $\eta\u{conv}$ is the efficiency of the QFC. 
From the relations, 
\begin{eqnarray*}
\g2\u{s,ast}&=&
\frac{\normal{n\u{ast}n\u{s}}}{\expect{n\u{s}}\expect{n\u{ast}}}\\
&=&\frac{\g2\u{s,asv} \zeta + 1}{\zeta + 1}
\end{eqnarray*}
is derived, 
which is Eq.~(\ref{eq:g2cross}) in the main text. 
We also obtain 
\begin{eqnarray}
 g\u{s,s|ast}^{(2)}
&=&
\frac{\normal{n\u{s}^2n\u{ast}}\expect{n\u{ast}}}{\normal{n\u{s}n\u{ast}}^2}
\nonumber\\
&=&
\frac{\normal{n\u{s}^2 n\u{asv}}+\normal{n\u{s}^2}\expect{n\u{noise}}}
{(g\u{s,ast}^{(2)})^2 \expect{n\u{s}}^2(\expect{n\u{asv}}+\expect{n\u{noise}})}
\nonumber\\
&=&
\frac{g\u{s,s|asv}^{(2)}(g\u{s,asv}^{(2)})^2\zeta +g\u{s,s}^{(2)}}
{(g\u{s,ast}^{(2)})^2 (\zeta + 1)}\nonumber\\
&=&
 \g2\u{s,s|asv}\left(
\frac{\g2\u{s,asv}}{\g2\u{s,ast}}\right)^2
\frac{\zeta}{\zeta +1}
+ \g2\u{s,s}
\left(
\frac{1}{\g2\u{s,ast}}
\right)^2
\frac{1}{\zeta +1},\nonumber
\end{eqnarray}
which is Eq.~(\ref{eq:SM}) in the main text. 

\section{Estimation of the collection efficiency}
\label{appD}

We roughly estimate the overall transmittance of 
the optical circuit of the AS photons. 
For this, we assume that the photons in modes S and AS 
generated from the atomic ensemble were initially in the two-mode squeezed state, 
which does not conflict with the observed values of 
$\g2\u{s,s}=1.58\pm 0.03$ and $\g2\u{asv,asv}=1.99\pm 0.03$. 
We define 
the overall transmittance of the optical circuit 
including the collection efficiency of the photons in mode S(AS), 
the transmittance of the frequency filter 
and the joint quantum efficiency of $\U{D}\u{s(v)1}$ 
and $\U{D}\u{s(v)2}$ by $\eta\u{s(asv)}$. 
By using the excitation probability $p\u{ex}$, 
$p\u{s}=p\u{ex}\eta\u{s}$, 
$p\u{v}=p\u{ex}\eta\u{asv}$ and 
$p\u{s,v}=p\u{ex}\eta\u{s}\eta\u{asv}$ are satisfied. 
From the equations and observed counts, 
we estimated $\eta\u{asv}=0.007$. 
Assuming that 
the quantum efficiency $0.6$ of the APD 
and the transmittance $0.25$ of the filter, 
the collection probability of the AS photons is about $0.05$. 
The other parameters were estimated to be 
$\eta\u{s}=0.006$ and $p\u{ex}=0.1$. 
We similarly estimated the overall transmittance $\eta\u{ast}$ 
of the circuit for the converted 1522-nm photons detected by $\Dtab$ 
to be $\eta\u{ast}=0.003$ which includes the efficiency of the QFC. 

\section{Estimation of the loss of the heralded AS photons}
\label{appE}

We derive the polarization ratio of the AS photons 
heralded by the V-polarized S photons. 
For this, we focus on a quantum state of a single atom. 
By the clean pulse, the atom is initially prepared in $g\u{a}$~($F=2$). 
We denote the atomic state in $g\u{a}$ with magnetic sublevel $m_F$ 
by $\ket{m_F}$. 
We first consider a case 
where the H-polarized AS photons and the V-polarized S photons are detected, 
which contributes the coincidence detection 
between modes S and AS in our experiment. 
Since we use the V-polarized write pulse and the H-polarized read pulse, 
the transition matrix 
with respect to the ordered basis $\{ \ket{m_F}\}_{m_F=2,\ldots,-2}$ 
from the initial state to the final state in $g\u{a}$ is described by 
$X\u{H}=X_{2,2'}X^+_{2',1}X^+_{1,2'}X_{2',2}$, where 
\begin{eqnarray}
 X_{2',2}=
\begin{bmatrix}
0& \sqrt{\frac{1}{12}} & 0 & 0 & 0\\
\sqrt{\frac{1}{12}} & 0 &  \sqrt{\frac{1}{8}} & 0 & 0\\
0 & \sqrt{\frac{1}{8}} & 0 &  \sqrt{\frac{1}{8}} & 0\\
0 & 0 & \sqrt{\frac{1}{8}} & 0 & \sqrt{\frac{1}{12}} \\
0 & 0 & 0 & \sqrt{\frac{1}{12}} & 0
\end{bmatrix}, 
\end{eqnarray}
\begin{eqnarray}
 X^{+}_{1,2'}=
\begin{bmatrix}
\sqrt{\frac{1}{4}} & 0 & \sqrt{\frac{1}{24}} & 0 & 0\\
0 & \sqrt{\frac{1}{8}} & 0 & \sqrt{\frac{1}{8}} & 0\\
0 & 0 & \sqrt{\frac{1}{24}} & 0 & \sqrt{\frac{1}{4}} 
\end{bmatrix}, 
\end{eqnarray}
$X^{+}_{2',1}=(X^{+}_{1,2'})^T$ and $X_{2, 2'}=X_{2', 2}$, 
up to a constant factor. 
The matrix elements are referred from 
the hyperfine dipole matrix elements 
for $\sigma^\pm$ transitions~\cite{Steck2001}. 
Next, we consider a case where 
the V-polarized AS photons and the V-polarized S photons are postselected, 
which does not contribute the coincidence detection between modes S and AS 
in our experiment. 
In this case, the transition matrix 
with respect to the ordered basis $\{ \ket{m_F}\}_{m_F=2,\ldots,-2}$ 
from the initial state to the final state is described by 
$X\u{V}=X_{2,2'}X^+_{2',1}X^-_{1,2'}X_{2',2}$, where 
\begin{eqnarray}
 X^{-}_{1,2'}=
\begin{bmatrix}
\sqrt{\frac{1}{4}} & 0 & - \sqrt{\frac{1}{24}} & 0 & 0\\
0 & \sqrt{\frac{1}{8}} & 0 & - \sqrt{\frac{1}{8}} & 0\\
0 & 0 & \sqrt{\frac{1}{24}} & 0 & - \sqrt{\frac{1}{4}} 
\end{bmatrix}, 
\end{eqnarray}
under the same constant factor. 

We assume that the magnetic sublevel of the initial atomic state in $g\u{a}$ 
is maximally randomized. 
In this case, the ratio of the amount of the H-polarized AS photons 
to that of the V-polarized AS photons is given by the 
ratio of $\U{tr}(X\u{H}^\dagger X\u{H})$ to 
$\U{tr}(X\u{V}^\dagger X\u{V})$. 
The value is calculated to be $33/8$. 
Thus, in our experimental setup, 
the loss of the AS photons heralded by the V-polarized S photons 
is estimated to be $8/41\sim 0.2$. 

\end{document}